\documentclass[11pt]{article}
\usepackage{mons2}
\input psfig.tex

\lefthead{Zerbi {\it et al}} \righthead{Observing $\gamma$ Dors
with MONS}

\begin{document}

\title{Understanding the $\gamma$ Doradus Phenomenon using MONS}

\author{F.M. Zerbi$^1$, C. Aerts$^2$, G. Handler$^3$, A.B. Kaye$^4$ and
E. Poretti$^1$ }

\affil{$^1$Osservatorio Astronomico di Brera, Via Bianchi 46,
23807 Merate (Lc), Italy \\ $^2$  Instituut voor Sterrenkunde,
Katholieke Universiteit Leuven, \\ Celestijnenlaan 200 B, B-3001
Leuven, Belgium\\ $^3$ SAAO, PO Box 9, Observatory 7935, South
Africa \\ $^4$ Los Alamos National Laboratory, X-TA, MS B-220, Los
Alamos NM 87545, USA}

\begin{abstract}
We discuss the results expected from observing $\gamma$ Dor stars with
the MONS satellite. We first describe how MONS Star Trackers will
consistently improve the present knowledge of this new class of variables.
We then discuss how the $\gamma$ Dor can be considered possible "bridges"
between opacity driven pulsations and stochastically excited solar like
oscillations. For this reason the possible inclusion of a $\gamma$ Dor
representative in the MONS main target list is suggested and discussed.
Three possible candidates are presented: $\gamma$ Dor, 9 Aur and BS 2740.
\end{abstract}

\section{A Brief Introduction to $\gamma$ Doradus Stars}

The $\gamma$ Dor stars have been recently defined (Kaye et al. 1999)  as a
class of variable stars in the lower-red part of the Cepheid instability
strip, undergoing $g$-mode pulsation.  Stars in this group show
(multi)periodic photometric and spectroscopic variability with typical
time scales between 0.3 and 3\,d and typical amplitudes below $0^m.1$. The
definition of the class was based on 12 objects contained in a so-called
{\it master list}. The principal physical properties of these stars are
reported in Table 1. Several objects suspected to belong to the class have
been recently added by Aerts, Eyer \& Kestens (1998) and Handler (1999a)
upon examination of HIPPARCOS light curves. At the present date the
data-base of known $\gamma$ Dor stars and related objects (Handler,
private communication) contains 141 objects divided into {\it Bona Fide}
(12), {\it Prime} (46) and {\it Other} (83) candidates. Bona fide objects
in the master list are stars for which extensive photometric and
spectroscopic data prove that pulsations must be the cause for the
variability. Prime and Other $\gamma$ Dor candidates are stars for which
either a moderate or a limited amount of photometric and/or spectroscopic
data is available, allowing to suspect or speculate that nonradial
pulsation (NRP) is the cause of the observed variations.

\begin{table}[h]
\footnotesize
\begin{center}
\begin{tabular}{cccccccccccc}
\tableline Star &$b-y$ & $<V>$ & ST & $v \sin i$ & $\pi$ &
[$Me/H$] & $M_V$ & $L$ & $T_{\rm eff}$ & $R$ & $M$
\\

& (mag) &  (mag) & & (kms$^{-1}$) & (mas)  & & (mag) &$[L_\odot]$
& (K) & $[R_\odot]$ & $[M_\odot]$
\\

\tableline

HD 224945 & 0.192 & 6.93 & F0 {\sc v} &  55 & 16.92 & $-0.30$ &
3.07 & 5.1 & 7250 & 1.43 & 1.51
\\

$\gamma$ Dor & 0.201 & 4.25 & F0 {\sc v} &  62 & 49.26 & $-0.02$ &
2.72 & 7.0 & 7200 & 1.70 & 1.57 \\

9 Aur & 0.217 & 5.00 & F0 {\sc v} &  18 & 38.14 & $-0.19$ & 2.89 &
6.0 & 7100 & 1.62 & 1.52 \\

BS 2740 & 0.219 & 4.49 & F0 {\sc v} &  40 & 47.22 & $-0.15$ & 2.86
& 6.2 & 7100 & 1.64 & 1.53 \\

HD 62454 & 0.214 & 7.15 & F1 {\sc v} &  53 & 11.18 & ~~0.16 & 2.39
& 9.5 & 7125 & 2.02 & 1.66 \\

HD 68192 & 0.227 & 7.16 & F2 {\sc v} & 85 & 10.67 & ~~0.05 & 2.30
& 10.5 & 7000 & 2.20 & 1.71 \\

HD 108100 & 0.234 & 7.14 & F2 {\sc v} & 68 & 12.10 &  $-0.03$ &
2.53 & 8.5 & 6950 & 2.01 & 1.62
\\

HD 164615 & 0.226 & 7.06 & F2 {\sc iv} & 66 & 14.36  & ~~0.20 &
2.82 & 6.5 & 7000 & 1.73 & 1.53 \\

BS 6767 & 0.183 & 6.40 & F0 {\sc v}n & 135 & 17.44  & $-0.10$ &
2.59 & 7.9 & 7300 & 1.76 & 1.61 \\

BS 8330 & 0.225 & 6.20 & F2 {\sc iv} & 38 & 19.90 & $-0.01$ & 2.67
& 7.4 & 7000 & 1.85 & 1.57 \\

BS 8799 & 0.181 & 5.99 & kA5 & 45 & 25.04 & $-0.36$ & 2.96 & 5.7 &
7375 & 1.46 & 1.54 \\

HD 224638 & 0.198 & 7.49 & F1 {\sc v}s & 24 & 12.56 & $-0.15$ &
2.98 & 5.5 & 7200 & 1.51 & 1.52 \\

\tableline \normalsize
\end{tabular}
\end{center}
\vskip-0.8truecm \caption{Observational parameters of the
confirmed $\gamma$ Dor variables and calculated or inferred
basic properties. Table taken from Kaye et al. (1999).
\label{deftab1} }
\end{table}

The most relevant problems in the observation of $\gamma$ Dor stars
concern the amplitudes and time scales of their variability. Amplitudes
are small and the clustering of the typical frequencies around 1\,d$^{-1}$
creates conflicts with the 1\,d$^{-1}$ alias typical of ground-based
single-site observations and requires to organize multi-longitude
campaigns. At the same time even in these campaigns problems of alignment,
improper correction for atmospheric extinction and fluctuation in the
instrumental setup often introduce spurious low-frequency terms that have
the chance to be misinterpreted as signal. For this reason $\gamma$ Dor
stars will benefit as many other classes of variables of continuous
out-of-atmospheric coverage as the one provided by a space mission like
MONS.

The interested community has largely discussed if $\gamma$ Dor stars
have any relevance for asteroseismology. These stars pulsate in $g$-modes
that in some sense carry more information on the stellar interior than
$p$-modes, but the low number of modes excited and the intrinsic
uncertainties in the application of standard mode-identification
techniques seem to make a seismic interpretation rather difficult. The
$\gamma$ Dor stars give us however two additional possibilities for
seismological investigations which other classes of pulsating star cannot
provide.

Firstly, the domain of $\gamma$ Dor stars in the HR diagram partly
overlaps with that of the $\delta$ Scuti stars (Handler 1999a). It can
therefore be suspected that stars showing both types of pulsation could be
present. This will allow to constrain the physical properties of a star
tightly from the $\delta$ Scuti-type $p$ modes, which helps in identifying
the $\gamma$ Dor-type $g$ modes which can then be used to probe the
star's deep interior. Indeed, short-period light variations have already
been detected for two $\gamma$ Dor stars (see Handler 1999b for
the first discovery).

Secondly, Guzik et al. (2000) pointed out the major role played by
convection in $\gamma$ Dor pulsation theory. These authors
calculated nonadiabatic pulsation properties of evolutionary
models lying near the $\gamma$ Dor instability region in the H-R
diagram. The models considered have relatively deep envelope
convection zones. By using the Pesnell (1990) non-adiabatic
pulsation code these authors calculated the $l=0,1,2$ pulsation
frequencies and found unstable high order $g$-modes with
frequencies between 4 and 25 $\mu$Hz (periods $\sim$ 0.4 to 3
days). Furthermore the mode kinetic energy was found to reach a
minimum at a frequency of about 11 $\mu$Hz (period of $\sim$ 1 d),
i.e. near the most commonly observed $\gamma$ Dor period. The
frequency spacing between modes with same $l$ and subsequent $n$
was found to be $\sim 0.1$ d$^{-1}$, exactly matching the typical
frequency spacing observed in multiperiodic $\gamma$ Dor stars.
These results strongly suggest that asteroseismology of $\gamma$
Dor stars might be easier as it seemed until very recently.

Another interesting outcome of the model calculations originates from the
"frozen-in convection" approximation used, in which fluctuations in the
convective luminosity are set to zero during the pulsation cycle. Because
convection does not adapt to transport the luminosity in this
approximation, the radiation is periodically blocked by the high opacity
at the convection zone base, resulting in pulsational driving. Therefore
the overstable $g$-modes in $\gamma$ Dor stars are first maintained by an
opacity bump at the basis of the convective layer which has nothing to do
with the classical $\kappa/\gamma$ mechanism and "tunnel" to the surface
through a convective envelope which is "lazy"  enough not to adjust and
damp them. Second the turbulent motion in the convective envelope could
trigger solar-like oscillations as in any other stars of this mass,
temperature and luminosity. Indeed the temperature and luminosity range
occupied by $\gamma$ Dor stars completely overlaps with the one in which
solar-like oscillations are predicted. In this sense $\gamma$ Dor stars
might represent a sort of "bridge" between overstable pulsation and
stochastically generated solar-like oscillations. The MONS primary
telescope represents an unique opportunity to investigate such a possible
link. For this reason we encourage the MONS science team to consider the
inclusion of an object of this kind in the primary target list.

\section{Observing $\gamma$ Doradus stars with the star trackers}

MONS star trackers will observe their targets continuously for about 30
days; this will greatly reduce the size of alias peaks. A computed MONS
spectral window predicts the strongest alias at .02 mHz ($\sim
1.7$\,d$^{-1}$) with an amplitude of $0.2\times 10^{-6}$. This number has
to be compared to the typical size of the 1\,d$^{-1}$ alias of
ground-based multisite campaigns for $\gamma$ Dor stars, which is of about
$3.1\times10^{-1}$ in amplitude or higher. Most of the available
literature on $\gamma$ Dor stars contains discussions of accurate
time-series analysis fighting against aliasing. In this sense the superb
quality of MONS spectral window will likely allow major improvements in
frequency recognition.

It has often been noticed in published frequency analyses (even among
$\gamma$ Dor candidates) that after subtraction of known signals from the
data suprisingly high residuals remained. In the best cases the mean
internal error achievable from the ground is 2000 ppm, on the average 4000
ppm is probably a more reliable value; this translates into noise levels
around 1000 ppm in the amplitude spectra in the frequency region of
interest. At the typical magnitudes of the proposed $\gamma$ Dor targets
MONS Star Trackers will allow clear 3$\sigma$ detections of signals with
amplitudes of the order of a few tens of ppm, providing information more
accurate by orders of magnitude with respect to the best observations from
Earth. In the case of {\it prime} or {\it other} candidates the suspected
presence of g-mode pulsation is mostly deduced from survey quality light
curves (see for example the left box of Figure 1). The residual scatter in
the light curves is typically a few percent ($\sim 10^4$ ppm). MONS STs
will improve the measurement accuracy by a factor of at least 100 allowing
to confirm the detection and to give an insight into the behaviour of the
star at the same time. Obviously, possibly superposed $\delta$~Scuti-type
pulsations would be easily detectable as well.

\section{Observation of $\gamma$ Doradus stars with the main CAM}

A star which is to be observed as a MONS prime target must be as bright as
possible and its $\gamma$ Dor nature must be well established. By
examination of Table 1 we can see that 3 of the bona fide $\gamma$ Dor
stars have apparent magnitudes between 4 and 5, i.e. in the typical
magnitude range of the solar-type variables included in MONS' prime
program. Those are $\gamma$ Dor, 9 Aur and BS 2740. The three stars have
been observed at least through one successful multi-longitude multi-colour
photometric campaign and the presence of NRP is established beyond any
reasonable doubt. The cases of $\gamma$ Dor and 9 Aur are particularly
interesting since these are the most thoroughly observed and studied
$\gamma$ Dor stars so far. We now turn to short descriptions of these
objects.

The bright star {\bf $\gamma$ Dor} was observed in the sixties during
survey work dedicated to some chemically peculiar A stars and was
suspected to show variations with an amplitude $\Delta V=0^m.05$. Cousins
(1992) observed it with high precision in 1983/84 (55 measurements in 120
nights). The analysis provided evidence of two close frequencies at 1.32
and 1.36\,d$^{-1}$, which was confirmed, together with their beat
frequency at 0.04\,d$^{-1}$, by a further intensive observing run in 1989.
The star has since then been object of two multi-longitude photometric and
one multi-longitude spectroscopic campaign (Balona et al. 1994, 1996). The
first campaign confirmed the two close frequencies at $f_1=1.32098$ and
$f_2$=1.36354\,d$^{-1}$ and provided evidence for the presence of a third
frequency at $f_3$=1.47447\,d$^{-1}$. The second campaign resulted in
clear evidence of line profile variations due to NRP.

\begin{figure}
\label{Ccurves} \psfig{figure=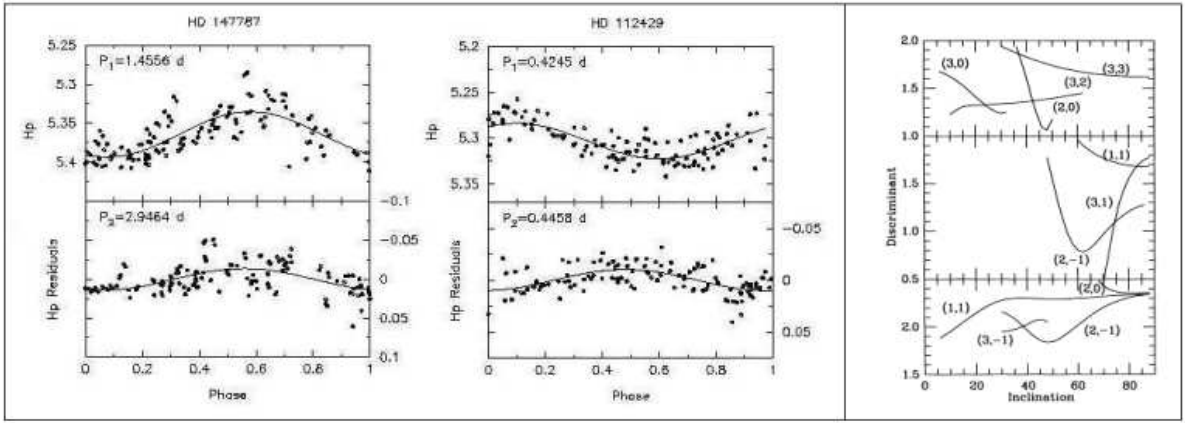,width=17cm}
\caption{Left box: phased Hipparcos light curves of suspected $\gamma$ Dor
stars reported by Aerts, Eyer \& Kestens (1998). Right box: results
of the moment analysis of $\gamma$ Dor presented by Balona et al.
(1996). The figure shows the discriminants in arbitrary units as a
function of the angle of inclination (in degrees) for various
modes ($l,m$). $f_1$, $f_2$ and $f_3$ are represented in the top,
mid and bottom panel, respectively.\label{monfig2}}
\end{figure}

The availability of line profiles sampled with sufficient resolution
allowed these authors to apply the {\it moment method} (Balona 1987, Aerts
et al. 1992) to this star. Such a method constructs a {\it discriminant},
the minimum value of which indicates the most suitable combination of
$i,l,m$ to fit the observed oscillation. The discriminant diagrams for
$\gamma$ Dor computed by Balona et al. (1996) are reported in
Figure~\ref{monfig2} (right box). Clearly the results are non-conclusive
since they do neither point to a unique solution nor, as they should, to a
unique value for the inclination. However, Balona et al. (1996) state that
the most probable angle of inclination for the star is $i=70^o$ and the
best estimate for the pulsational quantum numbers ($l, m$) are (3,3) for
$f_1$ and (1,1) both for $f_2$ and $f_3$.

Photometric variability of the bright northern star {\bf 9 Aurigae} was
reported by Krisciunas \& Guinan (1990). The star was then observed by
Krisciunas et al. (1995) through a multi-longitude campaign, however only
comprising sites in the continental US and Hawaii.  The analysis of these
data resulted in the detection of two frequencies at $f_1=0.794$ and
$f_2$=0.345\,d$^{-1}$.  Radial velocity measurements were also collected
during the campaign with the CORAVEL instrument in France.  Radial
velocity curves gave evidence for variability with frequency $f_2$, but no
sign of $f_1$, whereas the second output quantity of CORAVEL, the line
width, was dominated by $f_1$, as the photometry. This prompted Aerts and
Krisciunas (1996) to apply the moment method to the same CORAVEL data.
These authors reported a possible identification of the two frequencies as
the manifestation of an $l=3$, $|m|=1$ spheroidal mode and its toroidal
correction due to stellar rotation.

In order to overcome the aliasing problem, 9 Aur was re-scheduled for a
multi-longitude photometric campaign; the results were published by Zerbi
et al. (1997). This time a coverage of more than $18^h$ over $24^h$ was
achieved in a number of nights resulting in a substantial decrease of the
1\,d$^{-1}$ alias. The main frequency $f_1$ was confirmed and a new
frequency at $f_3$=0.7681\,d$^{-1}$ was discovered well above the level of
detection. However, the previously known mode at $f_2=0.3454$ could not be
clearly resolved. Its power was distributed into two peaks at
$f_{2a}$=0.2788d\,$^{-1}$ and at $f_1/2$=0.3975\,d$^{-1}$, the latter
causing the phased curve of the main signal to be double-wave-shaped. Most
of the measurements in this campaign were collected in the Str\"omgren
system. This made it possible to observe how the oscillations manifested
themselves in the $c_1$ index (tracing the luminosity) and the $(b-y)$
index (tracing the effective temperature) for the principal frequency. It
was found that they occurred in phase within the errors. This means that
geometrical displacements are not relevant for the oscillations, which can
be assumed as due basically to temperature variations. To some extent this
is what one would expect for a low-order $g$-mode oscillation.

The light variability of the F0 IV star {\bf BS 2740} was discovered by
Hensberge et al. (1981), who used it as a comparison star for Ap star
observations. On the basis of 20 $uvby$ measurements collected in 16
nights spanning 41 days, these authors derived a period of 0.936\,d$^{-1}$
and classified BS 2740 as a ``mild Ap star''. This evidence and the star's
position in the HR diagram combined with the detection of line-profile
variations in unpublished spectrograms prompted Poretti et al. (1997)  to
investigate its possible $\gamma$ Dor nature through a multi-longitude
campaign. The light curve of BS 2740 was explained with four frequencies,
$f_1=1.0434$, $f_2=0.9951$, $f_3=1.1088$ and $f_4$=0.9019\,d$^{-1}$ and
white noise. All the frequencies above are very close to 1\,d$^{-1}$,
which means that roughly the same phase of the oscillations would be seen
by a single-site observer. Multisite campaigns or space observations are
therefore required for this star.

\section{Conclusions}

The $\gamma$ Dor stars are known to exhibit $g$-mode pulsations at
time scales that make their observation difficult from the ground because
of aliasing problems. The location of the class in the HR diagram and some
theoretical arguments imply that solar-like oscillations could be excited
in these stars as well. The MONS satellite is then the ideal tool to
investigate both aspects taking profit of the clean spectral window of the
STs and the precision of the CAM. We therefore strongly support the
inclusion of a suitable candidate in the main target list and the
recording of as many targets as possible with the STs.

\end{document}


\title{Key aspects in the photometric monitoring of $\delta$ Sct stars}

\author{Ennio~Poretti}
\affil{Osservatorio Astronomico di Brera, Via E.~Bianchi 46, 
I-23807 Merate, Italy}



\begin{abstract}

The MONS experiment will be able to verify some of the
most important topics in the asteroseismologic studies
of \dsct\, stars.  This research will benefit not only
from the large increase in the number of detected frequencies,
but also from the possibility to  perform mode identification
by means of very precise two--colour photometry. To do that, an appropriate
frequency resolution, comparable to or better than what is currently
achieved by ground--based observations, is necessary and hence two
runs are recommended. To avoid over-scheduling of the Main
Camera, a possible combination of observations from the Main
Camera and from the Star Trackers is proposed. As a result, two key-objects
as FG Vir and BH Psc could be adequately monitored, affording the
possibility to enter into details of amplitude variations. As a 
by-product,
the same results could be obtained on two $\gamma$ Dor stars.

\end{abstract}

\section{Introduction}

In the last two decades a great observational effort has been undertaken
to solve the light curves of $\delta$ Sct stars. These complicated
variables, located on or just above the Zero--Age--Main--Sequence,
constitute the lower limit of the classical instability strip.
The presence of convective zones makes these stars
very interesting from the evolutionary point of view, since various
stages, from pre--main sequences to giant,  can be observed.

Photometric monitoring is the most practiced approach to study
their properties and several stars have been
deeply investigated; for  a review, see Poretti (2000).
However, the spectroscopic approach tells us that
many modes that are not photometrically detectable are actually excited
(Mantegazza 2000).

The observed frequencies are between 60 and 405 $\mu$Hz and
multiperiodicity is very common;
rotation acts as an important factor in the increase of the number of
excited modes.
 Only a few stars show a monoperiodic
behaviour above the current limit of the detectable amplitude from
ground, i.e. $\sim$1~mmag.
Of course, the number of detected modes is strongly
dependent on the accuracy, the number and the sampling of the measurements.

 The observed modes are in the domain
of pressure ($p$) modes; there is no observational evidence that
gravity ($g$) modes are excited in $\delta$ Sct stars, even if some cases
suggest they might be. At the moment, $g$-modes seem to be present only in
the $\gamma$ Dor stars, which in turn do not show $p$--modes.

How the photometric results can be used to identify modes (i.e. to
classify the oscillation in terms of quantum numbers $n$, $\ell$ and $m$)
is discussed  by Garrido (2000). 

\section {The state--of--art in the study of $\delta$ Sct stars}

Several approaches have been used in the past to attack the problem of the
detection of excited modes in their light curves. The debate on the
nature of the variations, multiperiodic or irregular, occupied the
years from 1974 to 1980. At that time the observations consisted
in time--series collected over a few nights, a practice 
that had been borrowed from the study of 
high amplitude \dsct\, stars. In the years that followed,
a number of teams observed more carefully \dsct\, stars 
and clearly evidenced their multiperiodicity, detecting stable
frequencies and obtaining very satisfactory fits. It appeared
also evident that nonradial modes are excited.
The number of frequencies is so high that the aliases at 1~cd$^{-1}$
become a serious problem in the mode detection. For that, multisite
campaigns were planned and the best results now available have been
obtained from these ground--based observations.

The effort made by the observers these last years allowed us to
detect a large number of frequencies in many stars: we will recall
here the cases of FG Vir, XX Pyx,
4 CVn and BH Psc. By means of multisite campaigns we are
able to detect terms with amplitude less than 1 mmag.
The mode identification techniques are
based on both the phase shifts and amplitude ratios of the light and
colour curves and the synergic approach performed by considering spectroscopic
curves. Their full exploitation should guarantee an important role for our
researches  in stellar physics.

Figure~1 shows the frequency content of three well-studied pulsators
in which all the terms with a half-amplitude larger than 1 mmag are
detected.
It appears that their content is completely different:
 high frequencies only
for XX Pyx, low frequencies only for 4 CVn, two groups of intermediate
 frequency values for FG Vir. 
The combination terms are also reported in the upper part of
each panel; 4 CVn displays both $f_i-f_j$ and $f_i+f_j$ terms, located 
right and left of the bunch of independent frequencies.
FG Vir displays only three combination modes in the same region of the
independent frequencies. XX Pyx displays only a 2$f$ harmonic, 
at 76.2~cd$^{-1}$.

The obvious conclusion is that $\delta$ Sct stars are
very complicated pulsators and that a general recipe cannot be applied 
to predict
the range of the excited modes or the importance of combination terms.
However, a curious similarity in the frequency content was noted by
Mantegazza \&
Poretti (1999) comparing 4 CVn to HD~2724: at least 7 frequencies have
almost the same value. 
\begin{figure}
\setlength{\textwidth}{4.0in}
\begin{center}
\plotone{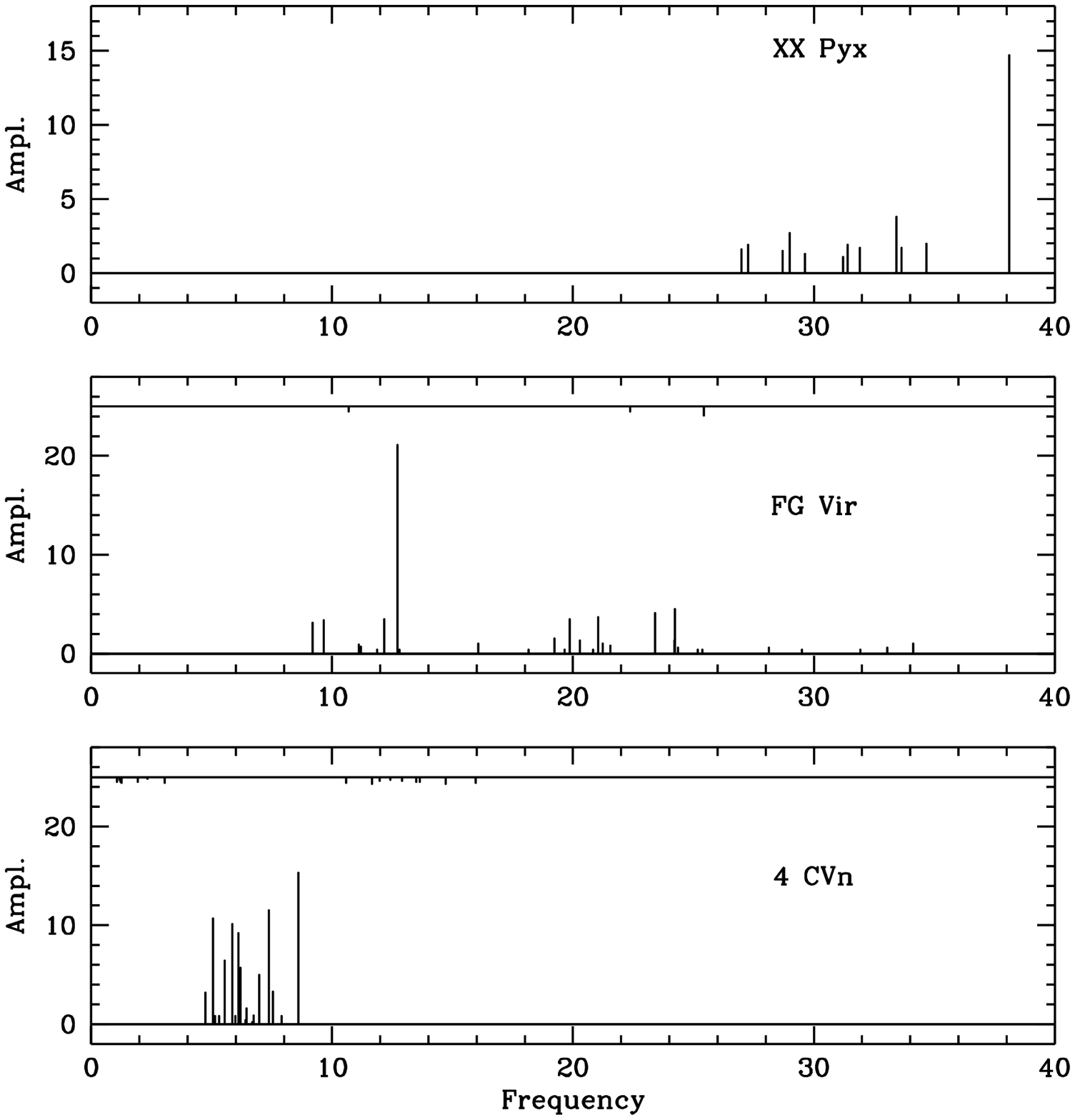}
\end{center}
\caption[]{\label{powsp} In $\delta$ Sct stars the different distributions
of the independent frequencies (lower line in each panel) and of the
combination terms (upper line in each panel; XX Pyx shows only a
harmonic term at 76.2~cd$^{-1}$, not displayed here)
are evident when comparing the
cases of XX Pyx, 4 CVn and FG Vir. Amplitudes are in mmag unit.}
\end{figure}

\section{The amplitude variations}

The variations of the amplitudes have been detected when a \dsct\,
star has been re--observed in order to investigate some unclear
facts: a comparison of the
mode amplitudes in the two different seasons often evidences strong
differences. Even if the data sampling can affect the reliability of
these conclusions, some results seem to be very much in favour of a physical
variability (Breger 2000).
These variations can be seen as a further complication, 
even if we can   
argue that the disappearance or the damping of some terms can enhance or
make discernible other modes, increasing the number of known frequencies.

Amplitude variations are observed in a large variety of stars, both
multiperiodic (XX Pyx, Handler et al. 1997; 4 CVn, Breger 2000;
44 Tau, Poretti et al. 1992) and monoperiodic (28 And, Rodr\'{\i}guez
et al.  1998; BF Phe, Poretti et al. 1996).
The observations of mode growth in the light curve of a relatively simple
pulsator as V663 Cas can clarify what happens in much more complicated
cases: some modes can be damped and then re--excited (Poretti et al.
1996). Some
observational facts favour this explanation even if the model of
a beating between two close frequencies with similar, constant amplitude
cannot be ruled out.
\smallskip

BH Psc is another example. It 
was observed by our group in 1991 (on two occasions), 1994 and 1995 at the
European Southern Observatory, Chile. The analysis of the first two
campaigns showed a very complex light variability resulting from
the superimposition of more than 10 pulsation modes with frequencies
between 5 and 12~cd$^{-1}$ and semi--amplitudes between 17 and 3 mmag
(Mantegazza, Poretti, \& Zerbi 1995).
The fit left a high r.m.s. residual 2.3 times greater than the one
measured between the two comparison stars.  A second photometric campaign
was carried out in October and November 1994; we hoped to reveal more
terms and to check the stability of their amplitudes. The analysis of
the new data allowed us to single out 13 frequencies (Mantegazza, Poretti,
\& Bossi 1996). They are concentrated
in the region 5--11.5~cd$^{-1}$: this distribution is slightly larger than 
the one observed for 4 CVn, but they can be considered very similar.
 However, more
terms should be present, with an amplitude below 3 mmag, since about 15\%
of the variance could not be explained with the detected terms and the noise.
Moreover, the standard deviation around the mean value was considerably
higher in 1991 than in 1994 (26 mmag against 18 mmag); considerations about
the amplitudes of the modes confirmed that the pulsation energy was
lower in 1994
than in 1991. Moreover, comparing the 1991 September and 1991
October data it is quite evident that we can detect such variations  on a
baseline of 1 month or less
(see Tab.~1, where the results of a preliminary analysis are
shown). 
\begin{table*}
\begin{center}
\caption[ ]{Preliminary values for the amplitudes of different terms 
observed in the light curve of BH Psc}
\smallskip
\begin{tabular}{rr rr rr}
\hline
\noalign{\smallskip}
\multicolumn{1}{c}{Frequency}& &\multicolumn{1}{c}{Sept. 1991}&
\multicolumn{1}{c}{Oct. 1991}&\multicolumn{1}{c}{Oct. 1994}&
\multicolumn{1}{c}{Oct. 1995}\\
\multicolumn{1}{c}{[cd$^{-1}$]}& &\multicolumn{4}{c}{[mmag]}\\
\noalign{\smallskip}
\hline
\noalign{\smallskip}
6.214 & & 17.5 & 17.1 & 15.0 & 18.4 \\
9.538 & & 13.8 &  9.8 &  1.1 & 8.2 \\
10.241& & 9.1 & 8.2 & 9.9 & 15.4 \\
10.561& & 5.7 & 16.5 & 2.3 & 14.1 \\
8.486 & & 13.3& 14.1&0.3 & 3.1 \\
7.118 & & 10.6 & 4.6 & 6.2 & 8.2 \\
8.788 & & 7.0 & 4.9 & 4.4 & 9.0 \\
11.423 & & 9.1 & 4.9 & 6.6 & 5.7 \\
\noalign{\smallskip}
\hline
\end{tabular}
\end{center}
\end{table*}

\section{Scientific topics to be matched by MONS}

The precision in the measurements achievable by MONS will of course
allow us to detect terms with amplitudes much smaller than 1 mmag.
This in turn will generate a much more detailed knowledge of
the frequency content  of
$\delta$ Sct variables. We could then obviously match one of the requests
 of the
asteroseismology theory, i.e. increasing the number of detected terms. 
This is quite obviuos. However, the values of the amplitudes of these
modes are also an important parameter that can help in the theoretical
models. The observational evidence is that these amplitudes are variable.
Can MONS then give us an important clue as regard this aspect ?

To reach the goal, observing twice the same field can be very useful.
Doing so,  we could:

\begin{enumerate}
\item gain in frequency resolution since, as shown in Fig.~1, there
are many close terms. It should also be noted that multisite campaigns 
span more than 50~d and that a single 30-50~d MONS run will not have a
sufficient frequency resolution to solve already known
close peaks. Owing to the expected
large number of peaks, interference between different terms can seriously
affect the quality of the results: an unappropriate frequency resolution
could be considered a weakness  of the MONS results;
\item  detect amplitude variations in a very clear way. The absence of
alias terms combined with the appropriate frequency resolution will allow
a clarification on the nature of the amplitude variations, i.e.
 beating or intrinsic variations.
\end{enumerate}
\section{A plausible solution}
In the current configuration, can MONS satisfy these two requirements~?

Among the $\delta$ Sct stars, FG Vir is by far the most studied: not
only an excellent observational basis is available, but also more and
more thorough theoretical models are proposed on the basis of the
observed frequencies (Breger et al. 1999). FG Vir
($\alpha$ 12$^{\rm h}14^{\rm m}15^{\rm s}$,
$\delta -5^{\circ}43^{\prime}00^{\prime\prime}$, 2000.0)
is located very close to $\beta$ Vir
($\alpha$ 11$^{\rm h}50^{\rm m}42^{\rm s}$,
$\delta +1^{\circ}45^{\prime}53^{\prime\prime}$, 2000.0), a F9~V star
candidate for searching solar--like oscillations. Then, FG Vir ($V$=6.1)
can be measured not only as a primary target with the Main Camera, but also
with the Forward--pointing star tracker (BST or SI-2) when $\beta$ Vir
is measured in the Main Camera. From this lucky coincidence, we
can obtain two runs on a well--studied \dsct\, star, greatly increasing
the number of detected frequencies and saving frequency resolution. 
Moreover, in both cases, it will be possible to measure II Vir, another
\dsct\ star close to FG Vir.

Is that the only result~? Certainly not. There would be one more
bonus. Looking through the backward--pointing
star tracker (FST or SI-1) we could see, in both cases, an interesting
field, located at 176$^{\circ}$ from FG and $\beta$ Vir.
Three intriguing objects are located in this field: namely, BH Psc
($\alpha$ 23$^{\rm h}59^{\rm m}31^{\rm s}$,
$\delta -2^{\circ}50^{\prime}37^{\prime\prime}$, 2000.0), i.e.
a $\delta$ Sct showing strong amplitude variations and  HD~224639 and
HD~224945, two $\gamma$ Dor variables.  The requirements about
frequency resolution and amplitude variations also apply to
$\gamma$ Dor variables.

As a conclusion, it looks possible for MONS to perform an observational
study of $\delta$ Sct stars in a very powerful way (increase in the
 number of detected terms, two--colour photometry, sufficiently long 
baseline allowing frequency resolution and amplitude monitoring)
just considering  some scheduling precautions and
putting an adequate target in the Main Camera.

\begin{table}
\begin{center}
\caption[ ]{A possible observing MONS schedule matching the scientific
requirements of the studies on $\delta$ Sct stars. See text for
explanations.}  
\begin{tabular}{lll lll lll }
\hline
\noalign{\smallskip}
\multicolumn{3}{c}{Main Camera}& 
\multicolumn{3}{c}{Frontward ST}&
\multicolumn{3}{c}{Backward ST}\\
\multicolumn{3}{c}{}& 
\multicolumn{3}{c}{(BST)}&
\multicolumn{3}{c}{(FST)}\\
\noalign{\smallskip}
\hline
\noalign{\smallskip}
  & FG Vir &  &    & $\beta$ Vir &  &    & BH Psc & \\
  &        &  &    & II Vir      &  &    & HD 224639 & \\
  &        &  &    &             &  &    & HD 224945 & \\
\hline
  & $\beta$ Vir & &      &  FG Vir & &      & BH Psc & \\
  &             & &      &  II Vir & &      & HD 224639\\
  &             & &      &         & &      & HD 224945\\
\noalign{\smallskip}

\noalign{\smallskip}
\hline
\end{tabular}
\end{center}
\end{table}